\documentclass[aps,prd,nofootinbib,superscriptaddress]{revtex4-2}

\usepackage{amssymb}

\usepackage{mathrsfs}         %%%%%%%%%%%% This is important. because we used Flower Fonts !!!!!
\usepackage{epsfig}
\usepackage{graphicx}
\usepackage{color}

\usepackage{amssymb}
\usepackage{fixltx2e}
\usepackage{bm}
\usepackage{amsmath,amsfonts,bm,dsfont}
\usepackage{verbatim}
\usepackage{mathrsfs}  %
\usepackage{graphicx}% Include figure files
\usepackage{subfigure}

\DeclareMathOperator{\Tr}{Tr}

\newcommand{\be}{\begin{eqnarray}}
	\newcommand{\ee}{\end{eqnarray}}
\makeatletter
\def\l@subsection#1#2{}
\def\l@subsubsection#1#2{}
\makeatother
\begin{document}

	\title{Multilayer Haldane model}% Force line breaks with \\

	\author{Xi Wu}
	\email{wuxi5949@gmail.com}
	\affiliation{Physics Department, Ariel University, Ariel 40700, Israel}
	
	\author{C.X. Zhang}
	\email{zhang12345s@sina.com}
	\affiliation{Physics Department, Ariel University, Ariel 40700, Israel}

	\author{M. A. Zubkov}
	
	\email{mikhailzu@ariel.ac.il}
	
	\affiliation{Physics Department, Ariel University, Ariel 40700, Israel}

	\begin{abstract}
		We propose the model of layered materials, in which each layer is described by the conventional Haldane model, while the inter - layer hopping parameter corresponds to the ABC stacking. We calculate the topological invariant $N_3$ for the resulting model, which is responsible for the conductivity of intrinsic quantum Hall effect. It has been shown that   in a certain range of the values of interlayer hopping parameter, the value of $N_3$ is equal to the number of layers multiplied by the topological invariant of each layer. At the same time this value may be calculated using the low energy effective theory.
	\end{abstract}
	
	% PACS, the Physics and Astronomy
	% Classification Scheme.
	%\keywords{Suggested keywords}%Use showkeys class option if keyword
	%display desired
	\maketitle
	
	%\tableofcontents

%%%%%%%%%%%%%%%%%%%%%%%%%%%%%%%%%%%%
\section{Introduction}
%%%%%%%%%%%%%%%%%%%%%%%%%%%%%%%%%%%%
Quantum Hall Effect (QHE) \cite{Klitzing:1980aa,Thouless:1982aa,Qi:2008aa} is one of the most remarkable phenomena in solid state physics. The quantization of Hall conductivity is so precise that it may be used as an etalon of its physical unit. The quantization is provided by the topological properties of matter. Originally the QHE has been considered in the presence of external magnetic field. Later the models of intrinsic anomalous QHE have been proposed, in which the QHE exists without external magnetic field. The very first model of this type is the so - called Haldane model  \cite{Haldane:1988aa}. It reveals correspondence with the tight - binding model of single - layer graphene \cite{KATSNELSON200720}. Several extra terms are added to the latter model that provide a non - trivial band topology resulted in the QHE. The Haldane model itself does not describe any real materials. However, qualitatively it describes the wide class of two - dimensional solids - state systems called now Chern insulators  \cite{Neupert:2011aa,Sheng_2011}.

The quantum Hall conductivity of intrinsic QHE may be expressed through the topological invariant $N_3$ composed of the two - point Green function
\cite{ISHIKAWA1987523,Golterman1992ub,Volovik_2009}. This invariant is also relevant for the description of topological phenomena in Helium-3 superfluid \cite{Volovik:1988aa}. As well as the ordinary QHE the intrinsic QHE may be both integer and fractional \cite{Neupert:2011aa,Sheng_2011}. In the majority of solid state systems the values of $N_3$ are $0$, $1$, $-1$. Larger values of $N_3$ are more rare. The systems with arbitrary values of the topological invariant have been considered, for example, in \cite{Sticlet:2012aa,Sticlet:2013aa}.

In the present paper we propose the model of two - dimensional layered systems based on the analogy to the multi - layer graphene \cite{Guinea:2006aa,McCann:2006aa,min2008electronic}. Various multi - layered systems are well - known in solid - state physics (see, for example, \cite{Trescher:2012aa} and references therein).
The model considered in our present paper may be called the multilayer Haldane model (with ABC stacking). We demonstrate that its value of $N_3$ is equal to the topological invariant of conventional (mono - layer) Haldane model multiplied by the number of layers, for a finite value of the inter - layer hopping parameter. As it was mentioned above, the Haldane model is defined on the hexagonal lattice. Therefore, our multi - layer Haldane Hamiltonian is related to the mono - layer Haldane Hamiltonian in the way similar to the models of multi - layer graphene \cite{Guinea:2006aa,McCann:2006aa,min2008electronic}. Notice that according to \cite{zhao2020tuning}, Chern insulators  are not practical for constructing devices if they have a single edge gapless mode because of large contact resistance. Correspondingly, it would be important to propose materials with large values of topological invariant $N_3$, which results in the large number of gapless edge modes.

 This paper is organized as following. In Sec (\ref{mlayerHald}) we present the n-layer Haldane model with ABC stacking; in Sec (\ref{topinv}) we calculate the topological invariant in two limits, the first in which the inter - layer hopping parameter vanishes and the second in which the inter - layer hopping parameter is much larger than the gap of the monolayer Haldane model; in Sec (\ref{Band}) we study the band structure and prove that the bands do not close for a finite value of inter - layer hopping: in Sec (\ref{Conc}) we make conclusions.

%%%%%%%%%%%%%%%%%%%%%%%%%%%%%%%%%%%%%%%%%%%%%%%%%%%%%%%%%%%%%%%%
\section{The multi - layer Haldane model}
%%%%%%%%%%%%%%%%%%%%%%%%%%%%%%%%%%%%%%%%%%%%%%%%%%%%%%%%%%%%%%%%
\label{mlayerHald}
In the case of ABC stacking, the multi - layer Haldane Hamiltonian is given by
\be\label{Hn}
	\mathbb{H}_{n}=\left[\begin{array}{ccccc}
	H_1 & \mathfrak{t}^{T} &  & &
	\\ \mathfrak{t} & H_1 & \mathfrak{t}^{T}  &  &
	\\  &  \mathfrak{t} & H_1 & \mathfrak{t}^{T} &
	\\  &  & &...  &
	\\ &  & & \mathfrak{t} &H_1\end{array}\right]_{2n\times2n}\,,
\ee
Here $\mathfrak{t}=\left[\begin{array}{cc}0 & t_{\bot} \\0 & 0\end{array}\right]$, while $t_\bot$ is the inter - layer hopping parameter.
$H_1$ is the Hamiltonian of mono - layer Haldane model
\begin{align}
	H_1=h_0 \sigma_0 + h_1 \sigma_1+h_2 \sigma_2+ h_3\sigma_3\,,
\end{align}
where
\begin{eqnarray}
	h_0 &=&  2t_2\cos\phi\sum_i\cos(\boldsymbol{p}\cdot \boldsymbol{b}_i)\\
h_1 &=& t_1\sum_i \cos(\boldsymbol{p}\cdot \boldsymbol{a}_i)\\
h_2 &=& t_1\sum_i \sin(\boldsymbol{p}\cdot \boldsymbol{a}_i)\\
h_3 &=& M-2t_2\sin\phi\sum_i\sin(\boldsymbol{p}\cdot \boldsymbol{b}_i).
\end{eqnarray}
The vectors $a_i$'s and $b_j$'s satisfy
\begin{align}
	&|\boldsymbol{a}_1|=|\boldsymbol{a}_2|=|\boldsymbol{a}_3|=a\nonumber
	\\
	&\cos  \langle \boldsymbol{a}_1,\boldsymbol{a}_2\rangle=\cos \langle \boldsymbol{a}_2,\boldsymbol{a}_3\rangle=\cos \langle \boldsymbol{a}_3,\boldsymbol{a}_1\rangle=-\frac{1}{2}\nonumber
	\\	
	&\boldsymbol{b}_i=\frac{1}{2}\epsilon_{ijk}( \boldsymbol{a}_j-\boldsymbol{a}_k)\,.
\end{align}
%%%%%%%%%%%%%%%%%%%%%%%%%%%%%%%%%%%%
We can also represent the Hamiltonian of the multi - layer system as follows
\begin{align}
	\mathbb{H}_{n} = h_0 \sigma_0+\sum_{i=1,2,3} h_i \sigma_i  + \hat{Y} t_\bot\,,
\end{align}
where

\be\label{Hn}
	\hat{Y}=\left[\begin{array}{ccccc}
	0& {y}^{T} &  & &
	\\ {y} & 0 & {y}^{T}  &  &
	\\  &  {y} & 0 & {y}^{T} &
	\\  &  & &...  &
	\\ &  & & {y} &0\end{array}\right]_{2n\times2n}\,,
\ee
with
$$
	y=\left[\begin{array}{cc}
	0& 1
	\\ 0 & 0 \end{array}\right]
$$
while by $\sigma_i$ we denote the "repeated" Pauli matrices:
\be\label{Hn}
	\sigma_i=\left[\begin{array}{ccccc}
	\sigma_i &  &  & &
	\\ & \sigma_i &   &  &
	\\  &  & \sigma_i &  &
	\\  &  & &...  &
	\\ &  & & &\sigma_i\end{array}\right]_{2n\times2n}\,,
\ee
Commutation relations follow:
\begin{eqnarray}
\{\sigma_i,\sigma_j\} &=& 2\delta_{ij}, \quad i,j = 0,1,2,3 \nonumber\\
\{\sigma_0,\sigma_j\} &=& 2\sigma_{j}, \quad j = 1,2,3 \nonumber\\
\{\sigma_3,\hat{Y}\} &=& 0\nonumber\\
\{\sigma_1,\hat{Y}\} &=& \Sigma_1\nonumber\\
\{\sigma_2,\hat{Y}\} &=& \Sigma_2 \nonumber\\
\{\sigma_0,\hat{Y}\} &=& 2 \hat{Y} \nonumber\\
\hat{Y}^2 &=& 1_\pm
\end{eqnarray}
where
\be\label{Hn}
	1_\pm=\left[\begin{array}{ccccc}
	1_-&  &  & &
	\\  & 1 &   &  &
	\\  &   & 1 &  &
	\\  &  & &...  &
	\\ &  & &  &1_+\end{array}\right]_{2n\times2n}\,,
\ee
with
$$
	1_-=\left[\begin{array}{cc}
	0& 0
	\\ 0 & 1 \end{array}\right]
$$
$$
	1_+=\left[\begin{array}{cc}
	1& 0
	\\ 0 & 0 \end{array}\right]
$$
while
\be\label{Hn}
	\Sigma_1=\left[\begin{array}{ccccc}
	0& 1 &  & &
	\\ 1 & 0 & 1  &  &
	\\  &  1 & 0 & 1 &
	\\  &  & &...  &
	\\ &  & & 1 &0\end{array}\right]_{2n\times2n}\,,
\ee
and
\be\label{Hn}
	\Sigma_2=\left[\begin{array}{ccccc}
	0& -i &  & &
	\\ i & 0 & -i  &  &
	\\  &  i & 0 & -i &
	\\  &  & &...  &
	\\ &  & & i &0\end{array}\right]_{2n\times2n}\,.
\ee
%%%%%%%%%%%%%%%%%%%%%%%%%%%%%%%%%%%%%%%%%%%%%%%%%%%%%%%%%%%%%%%%%%%
\section{Topological invariant for Hall conductivity}
%%%%%%%%%%%%%%%%%%%%%%%%%%%%%%%%%%%%%%%%%%%%%%%%%%%%%%%%%%%%%%%%%%%
\label{topinv}
The topological invariant \cite{ISHIKAWA1987523,Golterman1992ub,Volovik_2009} responsible for the conductivity of intrinsic QHE, is defined as follows
\be\label{TIN}
	\mathcal{N}[\mathbb{G}]=\frac{1}{3!}\int \frac{d^3 p}{(2\pi)^2} \,\epsilon^{ijk}\Tr(\mathbb{G}\partial_i \mathbb{G}^{-1}\mathbb{G}\partial_j \mathbb{G}^{-1}\mathbb{G}\partial_k \mathbb{G}^{-1})\,.
\ee
%Next we will calculate the topological invariant in the bilayer case and get a formal procedure that can be generalized into the case with n layers. In the last we will calculate the topological invariant in the general n-layer case.
% \subsection{Two theorems}
It is proportional to Hall conductivity: $\sigma_H =\frac{\mathcal{N}[\mathbb{G}]}{2\pi}$.  In these expressions $\mathbb{G}$ is the two - point Green function of electrons. This topological invariant defined through Green's function is equivalent to the usual Chern number description when interactions are absent.
Starting from Eq. (\ref{Hn}), the inverse of Green's function in multilayer Haldane model (with $n$ layers) can be written as
\be\label{Gn-1}
	\mathbb{G}_{}^{-1}=\left[\begin{array}{ccccc}
	Q & \mathfrak{t}^{T} &  & &
	\\ \mathfrak{t} & Q & \mathfrak{t}^{T}  &  &
	\\  &  \mathfrak{t} & Q & \mathfrak{t}^{T} &
	\\  &  & &...  &
	\\ &  & & \mathfrak{t} &Q\end{array}\right]_{2n\times2n}\,,
\ee
where $Q=i\omega-H_1$.
In this section, we consider two limits of the interlayer hopping $t_\bot$ when the value of topological invariant is easy to compute and found that they are the same. Since the band do not close for all finite $t_\bot$ (at least for the case ${\rm cos}\,\phi = 0$, see Sect. \ref{Band}),  we understand that the value of topological invariant remains the same when ${\rm cos}\,\phi $ does not deviate strongly from zero.
%%%%%%%%%%%%%%%%%%%%%%
\subsection{Interlayer hopping zero limit}
%%%%%%%%%%%%%%%%%%%%%%
First we consider the limit when the interlayer hopping is zero, namely $t_{\bot}=0$. Eq. (\ref{Gn-1}) becomes
\be
	\mathbb{G}_{}^{-1}=\left[\begin{array}{ccccc}
	Q & &  & &
	\\  & Q &   &  &
	\\  &   & Q &  &
	\\  &  & &...  &
	\\ &  & & &Q\end{array}\right]_{2n\times2n}\,. 	
\ee
The matrix $\mathbb{G}$ becomes a direct tensor product of the other matrices $Q$ . If a matrix $\mathbb{G}$ is a direct tensor product of the two other matrices $G_1$ and $G_2$, then the topological invariant, or the winding number, of $\mathbb{G}$ will be the sum of the topological invariant of $G_1$ and that of $G_2$.
 Namely, if $\mathbb{G}=\left[\begin{array}{cc} G_1 & 0 \\0 & G_2\end{array}\right]$,  then $N[\mathbb{G}]=N[G_1]+N[G_2]$  \cite{nakahara2003geometry}.

%From this theorem we understand that if the interlayer hopping parameter $t$ is zero, the value of the topological invariant of multilayer insulator is just the sum of that of each layer.

The proof is trivial. Namely, we have $\mathbb{G}^{-1}=\left[\begin{array}{cc} G^{-1}_1 & 0 \\0 & G^{-1}_2\end{array}\right]$,
\be
	\mathcal{N}[\mathbb{G}]&=&\frac{1}{3!}\int \frac{d^3 p}{(2\pi)^2}\,\epsilon^{ijk} \Tr(\mathbb{G}\partial_i \mathbb{G}^{-1}\mathbb{G}\partial_j \mathbb{G}^{-1}\mathbb{G}\partial_k \mathbb{G}^{-1})\nonumber \\
	&=&\frac{1}{3!}\int \frac{d^3 p}{(2\pi)^2}\, \epsilon^{ijk}\Tr(\left[\begin{array}{cc} G_1 & 0 \\0 & G_2\end{array}\right] \partial_i \left[\begin{array}{cc} G^{-1}_1 & 0 \\0 & G^{-1}_2\end{array}\right]\nonumber \\
	&&\left[\begin{array}{cc} G_1 & 0 \\0 & G_2\end{array}\right] \partial_j \left[\begin{array}{cc} G^{-1}_1 & 0 \\0 & G^{-1}_2\end{array}\right]\left[\begin{array}{cc} G_1 & 0 \\0 & G_2\end{array}\right] \partial_k \left[\begin{array}{cc} G^{-1}_1 & 0 \\0 & G^{-1}_2\end{array}\right])\nonumber \\
	&=&\frac{1}{3!}\int \frac{d^3 p}{(2\pi)^2}\, \epsilon^{ijk}\Tr(G_1\partial_i G^{-1}_1G_1\partial_j G^{-1}_1G_1\partial_k G^{-1}_1+G_2\partial_i G^{-1}_2G_2\partial_j G^{-1}_2G_2\partial_k G^{-1}_2)\nonumber \\
	&=&\mathcal{N}[G_1]+\mathcal{N}[G_2]\,.
\ee
Therefore,
\be
	\mathcal{N}[\mathbb{G}_{n}]=n\mathcal{N}[G]\,.
\ee
One can see that without the inter - layer hopping, the n-layer Haldane model would have the topological invariant with the value equal to the sum of the topological invariants of each layer. %Although above we considered the case, when all layers have the same Hamiltonian, this is true even when the Hamiltonians of the layers differ from each other.

%According to remark of the previous section when the value of $t_\bot$ is increased starting from zero, the value of the Hall conductivity remains the same if the Fermi energy is tuned in order to remain inside the gap. This value remains constant until the gap is closed (which may occur for a certain critical value of $t_\bot$).

%%%%%%%%%%%%%%%%%%%%%%%%
\subsection{Effective low energy theory}
%%%%%%%%%%%%%%%%%%%%%%%%
Let's consider the case when the interlayer hopping parameter is much larger than the gap of the monolayer Haldane model. This corresponds qualitatively to the real situation that takes place in multilayer graphene \cite{KATSNELSON200720}. In such a case, there is an effective description in which the topological invariant can be directly computed. At the would - be Fermi points Fermi point $K$ and $K'$ of graphene the off-diagonal parts of the Hamiltonian for the monolayer Haldane model vanish. In the small vicinity of $K$  we can write the Hamiltonian of monolayer Haldane model as (the similar expression will be at the $K'$ point):
\be
	H_1=\left[\begin{array}{cc}h_0 + h_3 & v\pi^{\dagger} \\v\pi & h_0 -h_3\end{array}\right]\,,
\ee
where $\pi=(p_1-K_1)+i(p_2-K_2)$,   and
$$h_3=M-2t_2\sin\phi\sum_i\sin(\boldsymbol{p}\cdot \boldsymbol{b}_i)\,,$$
 $$h_0=2t_2\cos\phi\sum_i\cos(\boldsymbol{p}\cdot \boldsymbol{b}_i)\,,$$
  $h_3$ is nonzero in a vicinity of $K$. The interlayer hopping parameter is much larger than the gap, which gives that $t_{\bot}>>|h_3|$ or $t_{\bot}>>t_2\,, M$.
 At this point we assume that the gap remains open in the given model, and the Fermi energy is inside the gap.

Let us show now that the effective $2\times2$ low energy Hamiltonian of the n-layer Haldane model has the form
\be\label{Hneff}
	H_{n}^{eff}=\left[\begin{array}{cc}h^{(n)}_3 & h^{(n)}(v\pi^{\dagger})^n \\ h^{(n)}(v\pi)^n & -h'^{(n)}_3\end{array}\right]\,,
\ee
using the mathematical induction similar to that of \cite{min2008electronic}. Here certain functions $h^{(n)}$, $h'^{(n)}_3$, $h^{(n)}_3$ depend on the size of the matrix.  First, indeed the monolayer Haldane model has such a form. Next, suppose that the n-layer Haldane model has the effective $2\times2$ low energy Hamiltonian with the form given by Eq. (\ref{Hneff}). Adding one more layer to the n-layer Haldane model, we obtain the Hamiltonian of the resulting model:
\be
	\left[\begin{array}{cccc}
	h^{(n)}_3 & h^{(n)}(v\pi^{\dagger})^n &  &
	\\h^{(n)}(v\pi)^n & -h'^{(n)}_3 & t_{\bot} &
	\\ & t_{\bot} & h_3^{(1)} & v\pi^{\dagger}
	\\ &  & v\pi & -h'^{(1)}_3
	\end{array}\right]\,.
\ee
The transformation based on the following matrix
\be
	S= \left[\begin{array}{cccc}
	1 &  &  &
	\\ &  &  & 1
	\\ &  & 1 &
	\\ & 1 &  &
	\end{array}\right]\nonumber
\ee
brings the Hamiltonian to the following form
\be\label{HPQ}
	\left[\begin{array}{cccc}
	h^{(n)}_3 &  &  &  h^{(n)}(v\pi^{\dagger})^n
	\\ & -h'^{(1)}_3 &v\pi  &
	\\ & v\pi^{\dagger} &  h_3^{(1)} & t_{\bot}
	\\h^{(n)}(v\pi)^n &  & t_{\bot} & -h'^{(n)}_3
	\end{array}\right]
	=\left[\begin{array}{cc}(H_{PP})_{2\times2} & (H_{PQ})_{2\times2} \\(H_{QP})_{2\times2} & (H_{QQ})_{2\times2}\end{array}\right]\,.
\ee
We can consider $H_{PQ}$ and $H_{QP}$ as perturbations in a vicinity of $\pi=0$, when $t_{\bot}>>|h_3|$. Then according to the degenerate state perturbation theory the effective $2\times2$ Hamiltonian is given by
\be\label{Heff}
	H^{eff}\approx H_{PP}-H_{PQ}\frac{1}{H_{QQ}}H_{QP}\,.
\ee
Applying Eq. (\ref{Heff}) to Eq. (\ref{HPQ}) we get the the effective $2\times2$ low energy Hamiltonian of the $(n+1)$ - layer Haldane model of the form of Eq. (\ref{Hneff}), in which functions $h^{(n)}$, $h'^{(n)}_3$, $h^{(n)}_3$ are defined by recursive relations:
\begin{eqnarray}
&& h^{(1)}_3 = h_3+h_0, \quad h^{(1)}=1, \quad h^{\prime (1)}_3=h_3-h_0\nonumber\\ &&
h^{(n+1)}_3 = h^{(n)}_3+h_0+\frac{|v\pi|^{2n}(h^{(n)})^2 (h_3+h_0)}{t_{\bot}^2+(h_3+h_0)(h_3^{\prime (n)}-h_0)} \nonumber\\&&
h^{(n+1)} = -\frac{t_{\bot}h^{(n)}}{t_{\bot}^2+(h_3+h_0)(h_3^{\prime (n)}-h_0)}\nonumber\\&&
h^{\prime(n+1)}_3 =  h_3-h_0+\frac{|v\pi|^{2}}{t_{\bot}^2+(h_3+h_0)(h_3^{\prime (n)}-h_0)}(h^{\prime (n)}_3-h_0)
\end{eqnarray}
  Recall that $t_\bot \gg h_3$. In the following we consider small vicinity of  $K$, where $|v\pi| \ll t_\bot$. One can see, that function $h^{(n)}$ in this vicinity is close to the constant $\frac{1}{(- t_\bot)^{n-1}}$, while $h^{(n)}_3$ is close to
  function $h_3+h_0$, which remains almost constant in this vicinity.

\subsection{Topological invariant for the low energy effective theory}
\begin{figure}[h]
	\centering  %
	\includegraphics[width=0.5\linewidth]{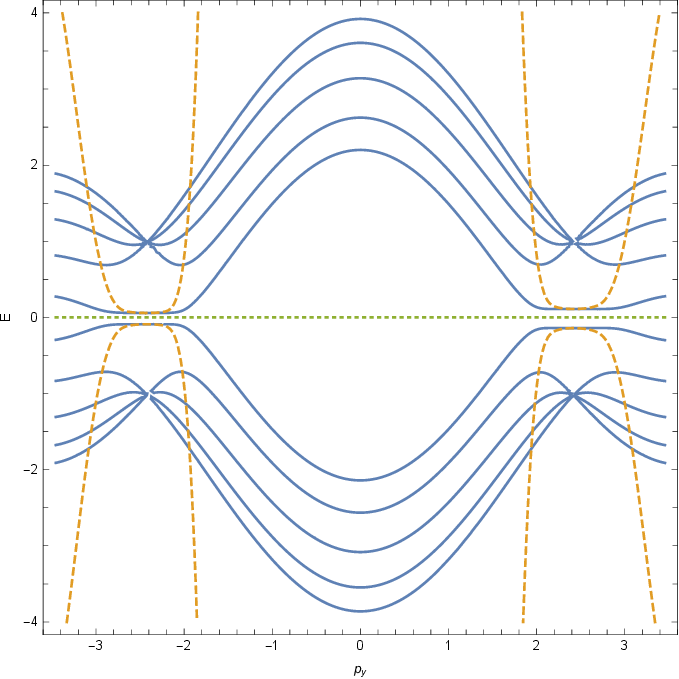}  %
	\caption{Dependence of energy on momentum $p_y$ at $p_x=0$ in the units of $t_1$ for the 5 - layer Haldane model at $2 \,t_2 \, {\rm cos}\,\phi = 2\, t_2 \, {\rm sin}\, \phi = 0.01\, t_1$, $M=0.1\, t_1$, $t_\bot = t_1$. The orange dashed line represents the energy bands of effective low energy theory. The green dotted line represents the line of zero energy $E=0$ counted from the Fermi level. }  %
	\label{fig2}   %
\end{figure}

If the gap is not closed when the value of $t_\bot$ is increased from zero to the given value, we may deform the Hamiltonian smoothly to the form, in which it is equal to
\be\label{H'ef}
	H_{n+1}'^{eff}&=&\left[\begin{array}{cc}h_0+h_3 & h^{(n+1)}(v\pi^{\dagger})^{n+1} \\ h^{(n+1)}(v\pi)^{n+1} & h_0-h_3\end{array}\right]\,.
\ee
It is possible to make this deformation in such a way that
 the poles of the Green function do not appear. Therefore, the given deformation cannot lead to modification of the topological number under consideration \cite{Sticlet:2012aa,wu2020anomalous}:
\be\label{NH}
	{N}[H]&=&\frac{1}{4\pi}\int d^2 \boldsymbol{p} \, \epsilon^{abc} \hat{h}_a \partial_{ p_x}\hat{h}_b\partial_{ p_y}\hat{h}_c
%	\\
%	&=& \frac{1}{4\pi}\int d^2 \boldsymbol{p} \, \epsilon^{abc} \frac{h_a}{|h|^3}\partial_{ p_x}h_b\partial_{p_y}h_c
	\,,
\ee
where
$$
\hat{h}_a = \frac{g_a}{\sqrt{\sum_a g_a^2}}
$$
and   $g_3 = h_3 $, $g_1 + i q_2 = h^{(n+1)}(v\pi^{\dagger})^{n+1}$, $h_0$ does not contribute.
From Eq.(\ref{NH}) we can understand that by stripping the same positive function from all the coefficients, we do not change the value of the  topological invariant.  Thus, we can use an even more simple Hamiltonian
\be
	H_{n+1}''^{eff}&=&\left[\begin{array}{cc}\frac{h_3}{|h^{(n+1)}|} & \text{sgn}(h^{(n+1)})(v\pi^{\dagger})^{n+1} \\ \text{sgn}(h^{(n+1)})(v\pi)^{n+1} & -\frac{h_3}{|h^{(n+1)}|}\end{array}\right]\,.	\label{Hef}
\ee
Moreover, the simultaneous change of the signs of $h_1$ and $h_2$ will not affect the topological invariant. Therefore,  we do not need to worry about the sign of $h^{(n+1)}$, and get a further simplified Hamiltonian with the same value of the topological invariant as that of Eq.(\ref{Hef})
\be
	H_{n+1}'''^{eff}&=&\left[\begin{array}{cc}\frac{h_3}{|h^{(n+1)}|} & (v\pi^{\dagger})^{n+1} \\(v\pi)^{n+1} & -\frac{h_3}{|h^{(n+1)}|}\end{array}\right]\,.	
\ee
This can already be put into Eq. (\ref{NH}) for the direct computation, i.e. we set $g_3 = \frac{h_3}{|h^{(n+1)}|}  $, $g_1 + i q_2 =  (v\pi^{\dagger})^{n+1}$. The result is
\be
	N[H^{eff}_{n+1}]=(n+1)N[H_1]\,.
\ee
The similar procedure can be applied to the vicinity of the $K'$ point. Finally, we obtain that the topological invariant (and thus the Hall conductivity) for the low energy effective theory is equal to the number of the layers times the value of topological invariant (the Hall conductivity) of  monolayer Haldane model. This pattern is illustrated by Fig. \ref{fig2}.

%%%%%%%%%%%%%%%%%%%%%%%%%%%%%%%%%%%%%
\section{Band structures}
%%%%%%%%%%%%%%%%%%%%%%%%%%%%%%%%%%%%%
\label{Band}
From Sect. \ref{mlayerHald}, the propagator of multi-layer Haldane model may be represented as
\begin{eqnarray}
\mathbb{G}_{n} &=& (({\cal E}-h_0) \sigma_0-\sum_{i=1,2,3} h_i \sigma_i  - \hat{Y} t_\bot\,)^{-1}\nonumber\\
&=& (({\cal E}-h_0) \sigma_0+\sum_{i=1,2,3} h_i \sigma_i  + \hat{Y} t_\bot\,)\Big( ({\cal E}-h_0)^2-\sum_{i=1,2,3} h^2_i   - t^2_\bot 1_\pm - t_\bot \Sigma_1 h_1 - t_\bot \Sigma_2 h_2 \Big)^{-1}
\end{eqnarray}
The energy levels $\cal E$ are given by solutions of equation
\if0\begin{eqnarray}
0 &=& {\rm det}\,  (({\cal E}-h_0) \sigma_0-\sum_{i=1,2,3} h_i \sigma_i  - \hat{Y} t_\bot\,) \nonumber\\
&=& {\rm det}\, \sigma_3 (({\cal E}-h_0) \sigma_0-\sum_{i=1,2,3} h_i \sigma_i  - \hat{Y} t_\bot\,)\sigma_3
 \nonumber\\
&=& {\rm det}\,  (({\cal E}-h_0) \sigma_0-h_3 \sigma_3 + \sum_{i=1,2} h_i \sigma_i  + \hat{Y} t_\bot\,)
\end{eqnarray}
It gives
\fi
\begin{eqnarray}\label{detmore}
 0 %&=& {\rm det}\,  (({\cal E}-h_0) \sigma_0-\sum_{i=1,2,3} h_i \sigma_i  - \hat{Y} t_\bot\,)  (({\cal E}-h_0) \sigma_0-h_3 \sigma_3 + \sum_{i=1,2} h_i \sigma_i  + \hat{Y} t_\bot\,)\nonumber\\
&=& {\rm det}\, \Big(({\cal E}-h_0)^2-\sum_{i=1,2,3} h^2_i   - t^2_\bot 1_\pm - t_\bot \Sigma_1 h_1 - t_\bot \Sigma_2 h_2   \Big)\nonumber\\
\end{eqnarray}
or
$$
0= {\rm det}_n({\cal E})
$$
where
\be\label{dnE}
	{\rm det}_n({\cal E})={\rm det}\,\left[\begin{array}{ccccc}
	F_-({\cal E}) & - t_\bot (h_1 - i h_2) & & &
	\\- t_\bot (h_1 + i h_2) & F({\cal E})  & - t_\bot (h_1 - i h_2)  &  &
	\\0 & - t_\bot (h_1 + i h_2)& F({\cal E})  & - t_\bot (h_1 - i h_2)&
	\\  &  & &...  &
	\\ &  & & - t_\bot (h_1 + i h_2) &F_+({\cal E}) \end{array}\right]_{2n\times2n}\,,
\ee
and
$$
F({\cal E})  = ({\cal E}-h_0)^2 - \sum_{i=1,2,3}h_i^2-t_\bot^2
$$
$$
F_+({\cal E})  = ({\cal E}-h_0)^2 - \sum_{i=1,2,3}h_i^2-t_\bot^2 1_+
$$
$$
F_-({\cal E})  = ({\cal E}-h_0)^2 - \sum_{i=1,2,3}h_i^2-t_\bot^2 1_-
$$
%There is a possibility of introducing more solutions in Eq. (\ref{detmore}) than that reality, 
From now on, we limit our discussion to the situation when $h_0=\text{constant}$ by setting $\cos \phi=0$. In such situation, the gap of ${\cal E}-h_0$ is the same as the gap of the energy ${\cal E}$ of the system.
Let us consider, for example, the particular case $n=2$. Then  $d_2({\cal E})$ is
\be\label{d2E}
	{\rm det}_2({\cal E})&=&{\rm det}\,\left[\begin{array}{cc}
	F_-({\cal E}) & - t_\bot (h_1 - i h_2)
	\\- t_\bot (h_1 + i h_2) &F_+({\cal E}) \end{array}\right]_{4\times4}
	\\
	&=&{\rm det}(F_-({\cal E})F_+({\cal E})- t_\bot ^2(h_1 ^2+ h_2^2) )\,.
\ee
\if0
\be\label{Hn}
{\rm det}\,\left[\begin{array}{ccccc}
	({\cal E}-h_0)^2 - \sum_{i=1,2,3}h_i^2 &0 &- t_\bot (h_1 - i h_2)& 0
	\\ 0& ({\cal E}-h_0)^2 - \sum_{i=1,2,3}h_i^2-t_\bot^2  & 0  &  - t_\bot (h_1 - i h_2)
	\\- t_\bot (h_1 + i h_2) & 0& ({\cal E}-h_0)^2 - \sum_{i=1,2,3}h_i^2-t_\bot^2 & 0
	\\ 0 &   - t_\bot (h_1 + i h_2)& 0 &({\cal E}-h_0)^2 - \sum_{i=1,2,3}h_i^2 \end{array}\right]\nonumber
\ee

Let us denote ${\cal F}({\cal E}) = ({\cal E}-h_0)^2 - \sum_{i=1,2,3}h_i^2 $.
We come to the following equation
\begin{eqnarray}
 0={\cal F}^2({\cal E})({\cal F}({\cal E}) -t_\bot^2)^2-2{\cal F}({\cal E})({\cal F}({\cal E})-t_\bot^2)t_\bot^2 (h_1^2+h_2^2) + t_\bot^4 (h_1^2+h_2^2)^2
\end{eqnarray}
\fi
The energy bands can be found analytically:
\begin{eqnarray}
{\cal E} &=& h_0 \pm \sqrt{\sum_{i=1,2,3} h_i^2 + \frac{t_\bot^2}{2} \Big(1 \pm \sqrt{1+4\frac{h_1^2+h_2^2}{t_\bot^2}\Big)}}
\end{eqnarray}
By observing that the inequality
\begin{eqnarray}
 t_\bot^2(\frac{h_1^2+h_2^2 + h_3^2}{t_\bot^2} + \frac{1}{2}) \ge \frac{t_\bot^2}{2} \sqrt{1+4\frac{h_1^2+h_2^2}{t_\bot^2}}
\end{eqnarray}
 always holds, and the “=" happens only when  $t_\bot \to \infty$ and  $h_3(p_1,p_2) = 0$.
One can see, that when the value of $t_\bot$ is increased from zero, the gap is not closed as far as $t_\bot$ is finite. However, it becomes infinitly close to zero at $t_\bot \to \infty$ at the points, where $h_3(p_1,p_2) = 0$. Such points do exist if the corresponding monolayer Haldane model has nonzero topological invariant. % We do not exclude, however, that for the larger number of layers the gap might be closed for a certain finite critical value $t^{(c)}_\bot$. But even so, there should obviously exist the region of the parameter $0< t_\bot < t_\bot^{(c)}$, where the gap remains open.
For $n>2$, it is difficult to get the explicit expression of energy bands, however, we can still show whether there is a gap closing: if $d_n({\cal E}=h_0)$ is nonzero always, we can say that there is no gap closing. And this is the case. In the following, we give one more examples of $n=3$  and then give a proof for generic $n$. The computation of Eq. (\ref{dnE}) is given in Appendix \ref{EBnlayer}.

For $n=3$
 \be\label{d3E}
	{\rm det}_3({\cal E})&=&{\rm det}\,\left[\begin{array}{ccc}
	F_-({\cal E}) & - t_\bot (h_1 - i h_2) &
	\\- t_\bot (h_1 + i h_2) & F({\cal E})  & - t_\bot (h_1 - i h_2)
	\\0 & - t_\bot (h_1 + i h_2)& F_+({\cal E}) \end{array}\right]_{6\times6}
	\\
	&=&{\rm det} (F_-({\cal E})F({\cal E}) F_+({\cal E})- t_\bot ^2(h_1 ^2+ h_2^2) (F_-({\cal E})+ F_+({\cal E})) \,,
\ee
and
\be
	{\rm det}_3({\cal E}=h_0)=h_3^2t_\bot^4+2h_3^2\sum_ih_i^2t_\bot^2+(\sum_ih_i^2)^3>0\,
\ee
which implies that the gap will remain open when $t_\bot < \infty$.
The energy spectra of 3-layered Haldane model are shown in Fig.\ref{fig.s}.

%%%%%%%%%%%%%%%%%%%%%%%%%%%%%%%%  2X2 figures %%%%%%%%%%%%%%%%%%%%%%%%%%%%%%%%
\if0
\begin{figure}[htbp]
\centering
\begin{minipage}[t]{0.48\textwidth}
\centering
\includegraphics[width=6cm]{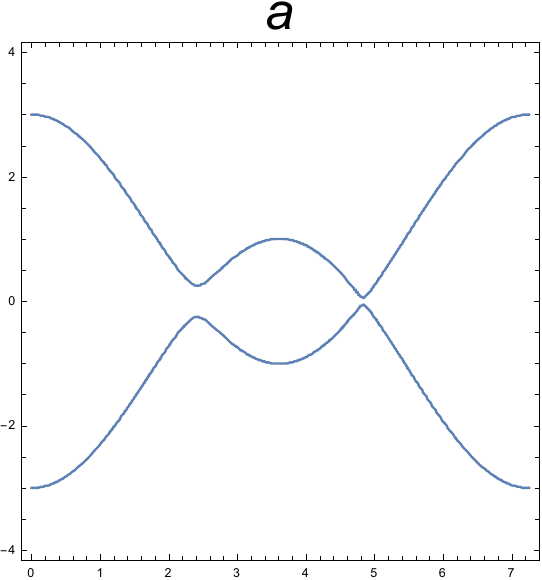}\nonumber
\caption{$t_\bot=0$}
\end{minipage}
\begin{minipage}[t]{0.48\textwidth}
\centering
\includegraphics[width=6cm]{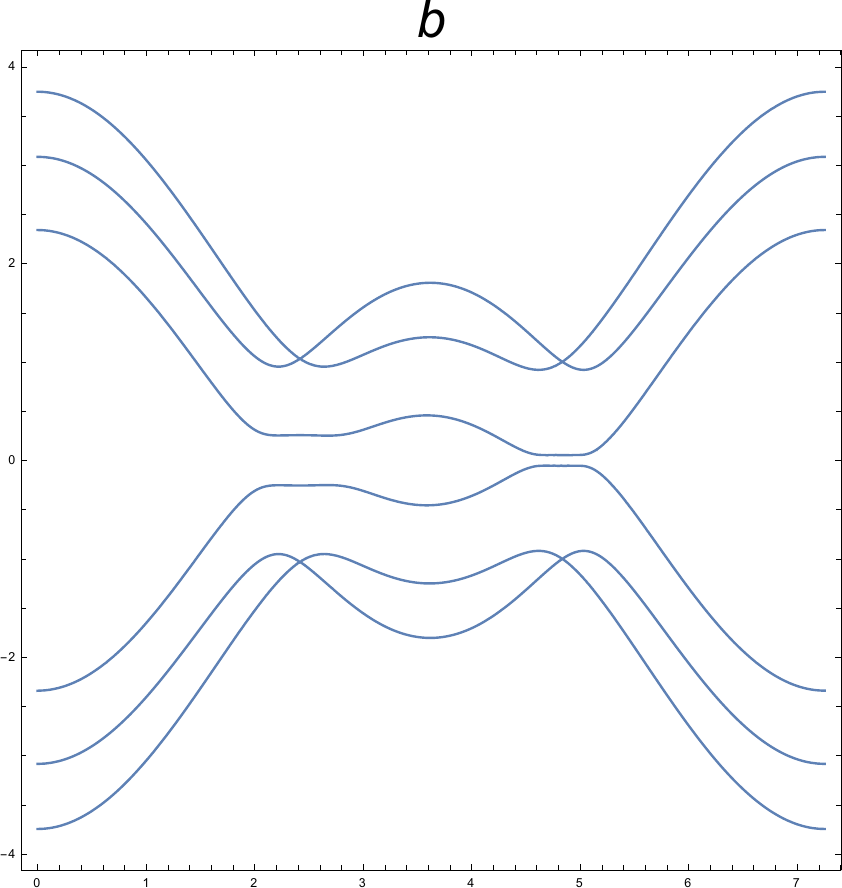}
\caption{$t_\bot=1$}
\end{minipage}
\quad

\vspace{1cm}
\centering
\begin{minipage}[t]{0.48\textwidth}
\centering
\includegraphics[width=6cm]{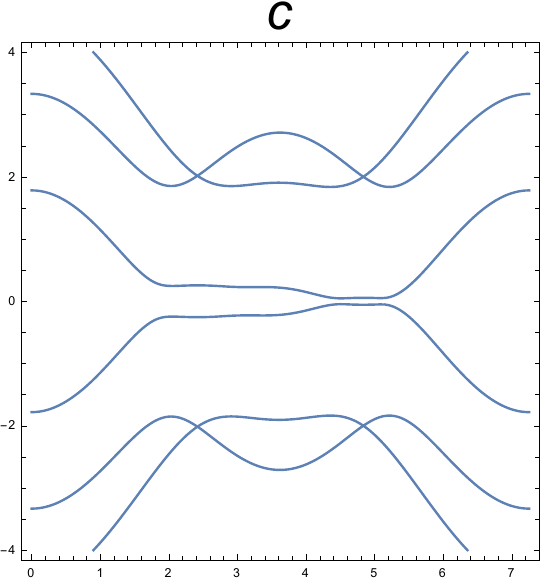}
\caption{$t_\bot=2$}
\end{minipage}
\begin{minipage}[t]{0.48\textwidth}
\centering
\includegraphics[width=6cm]{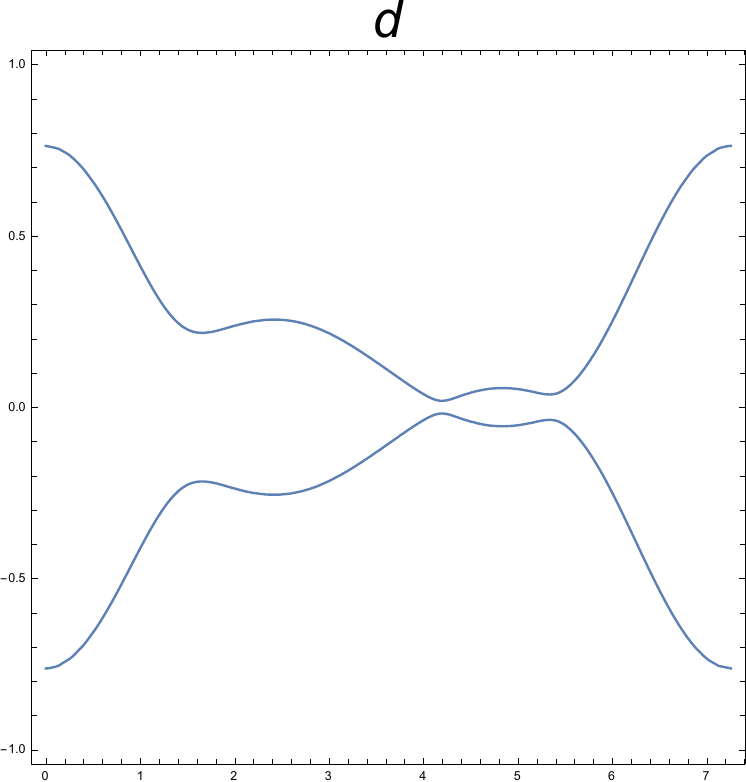}
\caption{$t_\bot=5$}
\end{minipage}
\caption{Energy spectra of three-layer Haldane model with various $t_{\bot}$ and $\cos \phi=0$.}
\label{fig.s}
\end{figure}
\fi
 \begin{figure}[t]
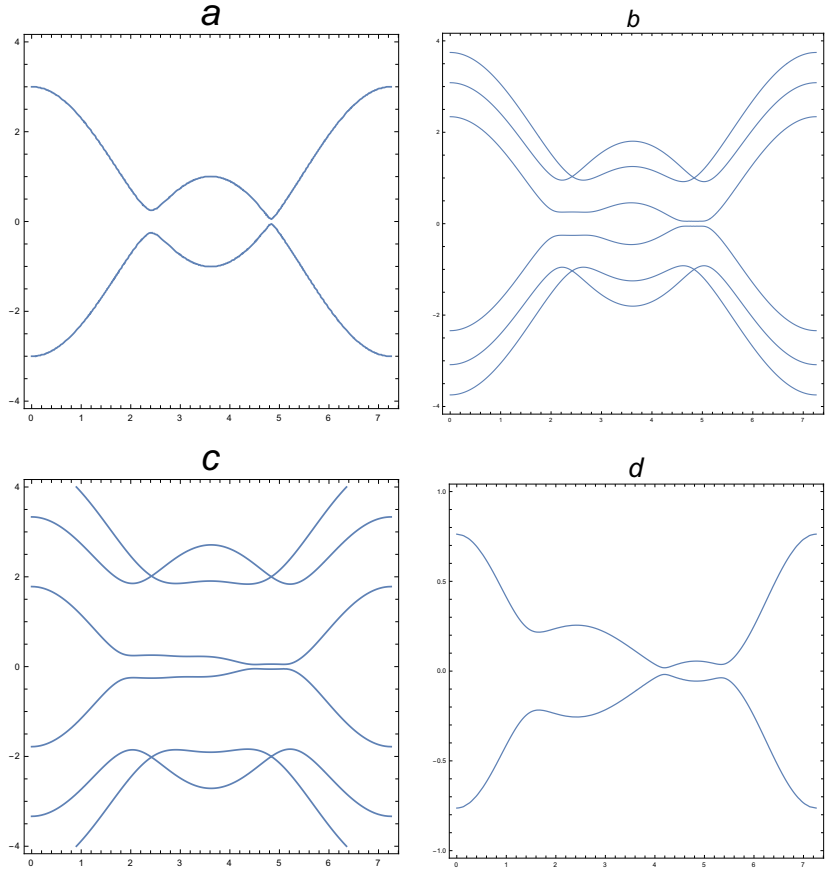

 \begin{center}
  \includegraphics[width=16em]{t=0.pdf} \quad
  \includegraphics[width=16em]{t=1.pdf} \\[1em]
  \includegraphics[width=16em]{t=2.pdf} \quad
  \includegraphics[width=16em]{t=5.pdf} %\\[1em]
 \end{center}
\caption{Energy spectra of three-layer Haldane model with $\cos \phi=0$, $t_1=1$, $M=0.1$, $t_2=0.03$. From $a$ to $d$, $t_{\bot}=0,1,2,5$ respectively.}% while in $g$, various values of $t_{\bot}$ are chosen.}
\label{fig.s}
\end{figure}
%%%%%%%%%%%%%%%%%%%%%%%%%%%%%%%%%%%%%%%%%%%%%%%%%%%%%%%%%%%%%%%%%%%

%For $n=4$, we have
%\be\label{d4E}
%	{\rm det}_4({\cal E})&=&{\rm det}\,\left[\begin{array}{cccc}
%	F_-({\cal E}) & - t_\bot (h_1 - i h_2) & &
%	\\- t_\bot (h_1 + i h_2) & F({\cal E})  & - t_\bot (h_1 - i h_2)  &
%	\\0 & - t_\bot (h_1 + i h_2)& F({\cal E})  & - t_\bot (h_1 - i h_2)
%	\\   &   & - t_\bot (h_1 + i h_2) &F_+({\cal E}) \end{array}\right]_{8\times8}
%	\\
%	&=& {\rm det} (F_-({\cal E})F^2({\cal E}) F_+({\cal E})- t_\bot ^2(h_1 ^2+ h_2^2) (F({\cal E})F_-({\cal E})+ F({\cal %E})F_+({\cal E})+F_-({\cal E})F_+({\cal E})) +t_\bot ^4(h_1 ^2+ h_2^2)^2 )
%	\,,
%\ee
%and
%\be
%	{\rm det}_4({\cal %E}=h_0)=h_3^2t_\bot^6+(2h_3^2\sum_ih_i^2+(\sum_ih_i^2)^2))t_\bot^4+3h_3^2(\sum_ih_i^2)^2t_\bot^2+(\sum_ih_i^2)^4>0\,.
%\ee

Now let us consider the case of generic $n$. For convenience we denote
\be
	A:=-F({\cal E}=h_0)>0\quad B:=-t_\bot(h_1-ih_2)\quad C:= -t_\bot(h_1+ih_2)\,.
\ee
  We find that from Eq. (\ref{dettilbD})
\be\label{dnEABC}
	2\sqrt{A^2-4BC}{\rm det}_n({\cal E}=h_0)
	&=&(A^2-t_\bot^2A-2BC)\Big ((\frac{A+\sqrt{A^2-4BC}}{2})^{n-1}- (\frac{A-\sqrt{A^2-4BC}}{2})^{n-1}\Big)
\nonumber\\
	&&+(A-t_\bot^2)\sqrt{A^2-4BC}\Big ((\frac{A+\sqrt{A^2-4BC}}{2})^{n-1}+(\frac{A-\sqrt{A^2-4BC}}{2})^{n-1}\Big)
\,.
\ee
We omitted the symbol ``$\det$" in the up equation since $A,~B,~C$ are proportional to the identity matrix.
Despite of its complicated form, we can prove that indeed ${\rm det}_n({\cal E}=h_0)>0$.
Denote
\be
	X:=A^2-t_\bot^2A-2BC, \quad Y:=(A-t_\bot^2)\sqrt{A^2-4BC}\,,
\ee
Eq. (\ref{dnEABC}) becomes
\be\label{dnEXY}
	2\sqrt{A^2-4BC}{\rm det}_n({\cal E}=h_0)
	&=&(Y+X)(\frac{A+\sqrt{A^2-4BC}}{2})^{n-1}+(Y-X) (\frac{A-\sqrt{A^2-4BC}}{2})^{n-1}\ge0
\ee
because
\be
	A+\sqrt{A^2-4BC}>0,\nonumber
	\\
	A-\sqrt{A^2-4BC}\ge0,\nonumber
	\\
	Y\ge\pm X,
\ee
because
\be
	Y^2-X^2=4t_\bot^4(h_1^2+h_2^2)h_3^2\ge0
\ee
and
\be
	Y=\sum_ih_i^2\sqrt{(\sum_ih_i^2-t_\bot^2)^2+4h_3^2t_\bot^2}\ge0\,.
\ee
Eq. (\ref{dnEXY}) tells us that
\be
	{\rm det}_n({\cal E}=h_0)\ge0\,.
\ee
Next we are going to show that the ``$=$" cannot hold so there is no band closing for finite $t_\bot$. The necessity condition for ${\rm det}_n({\cal E}=h_0)=0$ is, from Eq. (\ref{dnEXY}),
\be
	(Y+X)(\frac{A+\sqrt{A^2-4BC}}{2})^{n-1}+(Y-X) (\frac{A-\sqrt{A^2-4BC}}{2})^{n-1}=0\,,
\ee
which can be satisfied only when
\be
 \left\{ \begin{array}{rcl}\label{YXAB}
      &  Y+X=0 \\
       &A-\sqrt{A^2-4BC}=0
 \end{array}\right.
 \ee
or
\be
 \left\{ \begin{array}{rcl}\label{YXYX}
         Y+X=0 \\Y-X=0  \,.
 \end{array}\right.
 \ee
The two equations in Eq. (\ref{YXAB}) can not be satisfied simultaneously, while Eq. (\ref{YXYX}) give
\be
	A^2=4BC\,.
\ee
Considering the limit
\be
	A^2-4BC\to 0\,,
\ee
from Eq. (\ref{dnEABC}) we derive
\be
	{\rm det}_n({\cal E}=h_0)\to 2 \sum_i h_i^2 (\frac{\sum_ih_i^2+t_\bot^2}{2})^{n-1}>0\,.
\ee
So finally we conclude that for any finite value of $t_\bot$
\be
	{\rm det}_n({\cal E}=h_0)>0\,.
\ee
This means that the gap does not close for any $n$ layer Haldane model. Thus the value of topological invariant remains the same when we tune  $t_\bot$.
Our discussion in this section is based on the assumption $\cos \phi=0$. But this may be loosened by assuming the order of $\cos \phi$ is smaller than 1 so that $h_0$ is almost flat in the Brillouin zone comparing with the change of $h_1$ $h_2$ $h_3$.% Moreover, from Fig (\ref{fig.s}), we can see that the size of the gap seems not to depend on the value of $t_{\bot}$ in the 3-layer case, which means it is the same as the single-layer case, we conjecture that .

\begin{figure}[h]
	\centering  %
	\includegraphics[width=0.5\linewidth]{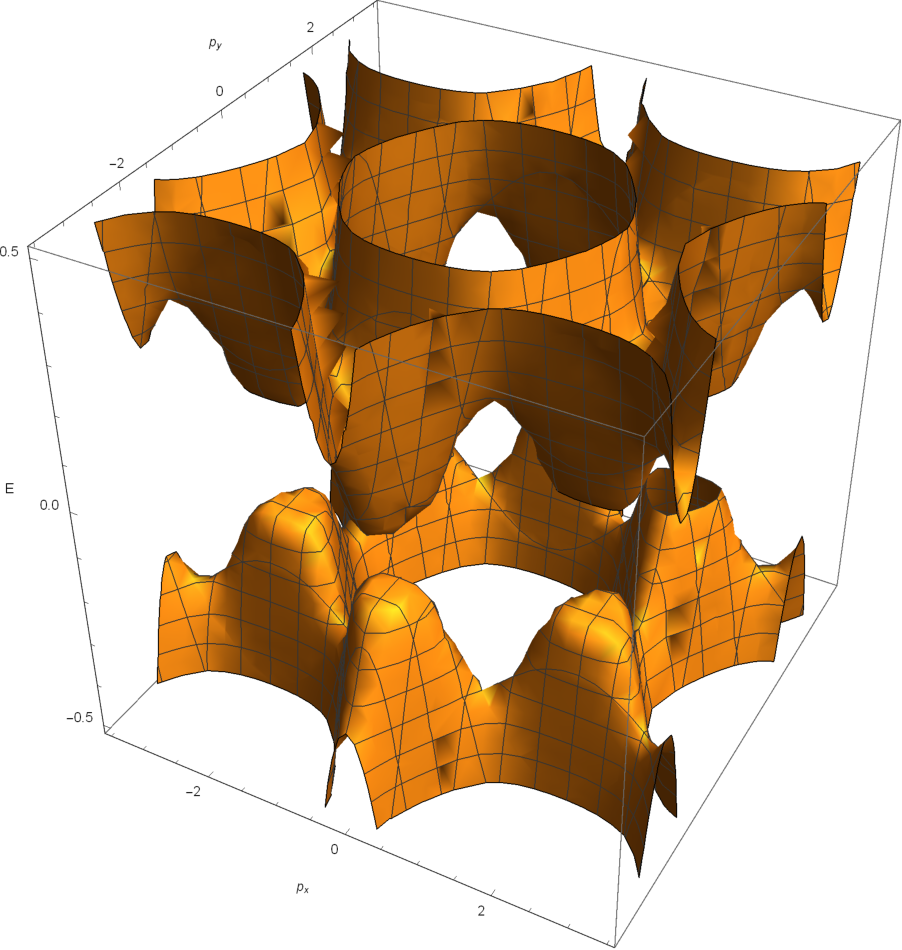}  %
	\caption{Dependence of energy on momenta in the units of $t_1$ (the lowest bands) for the 5 - layer Haldane model at $2\, t_2 \, {\rm cos}\,\phi = 2\, t_2 \, {\rm sin}\, \phi = 0.01\, t_1$, $M=0.1\, t_1$, $t_\bot = t_1$.}  %
	\label{fig1}   %
\end{figure}

%%%%%%%%%%%%%%%%%%%%%%%%
\section{Conclusions}
%%%%%%%%%%%%%%%%%%%%%%%%
\label{Conc}
In this paper, we have studied the topological invariant and the band structure responsible for the intrinsic QHE conductivity of the multilayer Haldane model with ABC stacking. We considered two limits of interlayer hopping when it is zero and when it is sufficiently large, given also by the effective low energy model containing only two energy bands. We showed that in both cases, the value of the topological invariant is equal to the number of layers times the value of the topological invariant of monolayer Haldane model for finite values of interlayer hopping parameter. Moreover we showed that the band do not close for arbitrary finite value of interlayer hopping when $h_0=0$ or almost constant.%Thus this is true for a finite value of interlayer hopping.
%In this paper, we have studied the band structure and the topological invariant responsible for the intrinsic QHE conductivity of the multilayer Haldane model with ABC stacking. We showed that the band do not close for arbitrary finite value of interlayer hopping. Moreover, we considered two limits of interlayer hopping when it is zero and when it is sufficiently large, given also by the effective low energy model containing only two energy bands. We showed that in both cases, the value of the topological invariant is equal to the number of layers times the value of the topological invariant of monolayer Haldane model for finite values of interlayer hopping parameter. Thus this is true for a ll  finite value of interlayer hopping.}

It is worth mentioning, that in our analysis we disregard completely the consideration of both inter - electron interactions and disorder. According to rather general considerations of \cite{ZZ_2019} the interactions cannot affect the value of Hall conductivity, at least, on the level of perturbation theory. The similar statement may also be drawn for the role of disorder \cite{ZZ_2019_2} as long as we consider the value of conductivity averaged over the sample area. In the latter case, however, the Hall current is expected to be concentrated along the boundary.

The multi - layered system qualitatively similar to the one discussed here, in principle, may be realized experimentally \cite{zhao2020tuning} via the appropriate crystal growth. Such a growth being performed would represent an engineering of the system with arbitrary large integer value of the topological invariant responsible for intrinsic anomalous quantum Hall effect.

\appendix
\section{Energy bands of n-layer Haldane model}
\label{EBnlayer}

Let's calculate Eq. (\ref{dnE}) for generic $n$. We do this in steps. First we solve the following determinant
\be\label{dn}
	\bold{d}_n={\rm det}\,\left[\begin{array}{ccccc}
	f & b & & &
	\\c & f  & b &  &
	\\0 & c & f & b&
	\\  &  & &...  &
	\\ &  & & c &f \end{array}\right]_{n\times n}\,,
\ee
where $f$, $b$ and $c$ are just numbers. From this definition we can derive that
\be\label{dndn-1}
	\bold{d}_n=f \bold{d}_{n-1}-bc\bold{d}_{n-2} \quad \mbox{for} ~ n\ge2
\ee
and
\be
	\bold{d}_1=f ~ \mbox{and} ~ \bold{d}_0=1\,.
\ee
Rewriting Eq. (\ref{dndn-1}) as
\be
	\bold{d}_n-r \bold{d}_{n-1}&=&(f-r) (\bold{d}_{n-1}-r \bold{d}_{n-2} ) \label{dxddxd}\,,
%	\\
%	&=&(f-x)^{n-1} (\bold{d}_{1}-x \bold{d}_{0} )
\ee
in which $r$ is to be determined, such that the above equation is equivalent to  Eq. (\ref{dndn-1}),
therefore we get an equation for $r$
\be
	r^2-fr+bc=0
\ee
which is solved by
\be\label{xpm}
	r_{\pm}=\frac{f\pm\sqrt{f^2-4bc}}{2}\,.
\ee
From Eq.(\ref{dxddxd}) we get
\be
 \left\{ \begin{array}{rcl}
         \bold{d}_n - r_+ \bold{d}_{n-1}=(f-r_+)^{n-1} (\bold{d}_{1}-r_+ \bold{d}_{0} )
                                      =r_-^{n-1} (\bold{d}_{1}-r_+ \bold{d}_{0} ) \\
         \bold{d}_n - r_- \bold{d}_{n-1} =(f-r_-)^{n-1} (\bold{d}_{1}-r_- \bold{d}_{0} )
                                      =r_+^{n-1} (\bold{d}_{1}-r_- \bold{d}_{0} )
 \end{array}\right.
 \ee
from which we get
\be\label{dnd1d0}
	\sqrt{f^2-4bc} \, \bold{d}_n
   &=& r_+^{n} (\bold{d}_{1}-r_- \bold{d}_{0})- r_-^{n} (\bold{d}_{1}-r_+ \bold{d}_{0}) \\
   %\Big((\frac{f+\sqrt{f^2-4bc}}{2})^n-(\frac{f-\sqrt{f^2-4bc}}{2})^n\Big)\bold{d}_1\nonumber\\
   %&&-\Big(\frac{f-\sqrt{f^2-4bc}}{2}(\frac{f+\sqrt{f^2-4bc}}{2})^n-\frac{f+\sqrt{f^2-4bc}}{2}(\frac{f-\sqrt{f^2-4bc}}{2})^n\Big)\nonumber
	%\\
	&=&(\bold{d}_1-\frac{f}{2}) \Big(r_+^n - r_-^n\Big)
	   +\frac{\sqrt{f^2-4bc}}{2}\Big(r_+^n + r_-^n\Big)
\ee
The above equation is valid, even if the value of $\bold{d}_1$ is some other number than $f$.

Next we consider a little more general case
\be\label{dn}
	\tilde{\bold{d}}_n={\rm det}\,\left[\begin{array}{ccccc}
	f_- & b & & &
	\\c & f  & b &  &
	\\0 & c & f & b&
	\\  &  & &...  &
	\\ &  & & c &f_+ \end{array}\right]_{n\times n}\,,
\ee
where the diagonal elements are all $f$ except the first $f_-$ and the last $f_+$.
Therefore, starting from the initial terms $d_1=f_+$ and $d_0=1$,
and following the above steps, terms $\tilde{\bold{d}}_{l}=\bold{d}_{l}$
with $2\le l \le n-1$ can be obtained.
From the relation $\tilde{\bold{d}}_n=f_-\tilde{\bold{d}}_{n-1}-bc\tilde{\bold{d}}_{n-2}
=\bold{d}_{n-1}-bc \bold{d}_{n-2}$,
and comparing with Eq.(\ref{dndn-1}), we obtain a
relation between $\tilde{\bold{d}}_n$ and $\bold{d}_{n-1}$, $\bold{d}_{n}$:
\be\label{itldndn-1dn}
	\tilde{\bold{d}}_n=f_-\bold{d}_{n-1}-bc\bold{d}_{n-2}=(f_--f)\bold{d}_{n-1}+\bold{d}_{n}\, ,
\ee
but with $\bold{d}_{1}=f_+$ here.
Substituting Eq. (\ref{dnd1d0}) into Eq. (\ref{itldndn-1dn}) we get
\be
	\sqrt{f^2-4bc}\tilde{\bold{d}}_n&=&(f_--f)\Big[(\bold{d}_1-\frac{f}{2}) \Big(r_+^{n-1}-r_-^{n-1}\Big)+\frac{\sqrt{f^2-4bc}}{2}\Big(r_+^{n-1}+r_-^{n-1}\Big)\Big]\nonumber
	\\
&&+(\bold{d}_1-\frac{f}{2}) \Big(r_+^n-r_-^n\Big)+\frac{\sqrt{f^2-4bc}}{2}\Big(r_+^n+r_-^n\Big)\,.
\ee
With some manipulations, we derive
\be\label{tildn}
	\sqrt{f^2-4bc}\,\tilde{\bold{d}}_n
&=&\Big((f_- -\frac{f}{2})(f_+ -\frac{f}{2}) +\frac{f^2-4bc}{4}\Big)\Big(r_+^{n-1}-r_-^{n-1}\Big)\nonumber
	\\
	&&+(f_-+f_+-f)\frac{\sqrt{f^2-4bc}}{2}\Big(r_+^{n-1}+r_-^{n-1}\Big)\,.
\ee

Next, we consider the case which is related with our n-layer model:
\be\label{Dn}
	\widetilde{\bold{D}}_n={\rm det}\,\left[\begin{array}{ccccc}
	F_- & B & & &
	\\C & F  & B &  &
	\\0 & C & F & B&
	\\  &  & &...  &
	\\ &  & & C &F_+ \end{array}\right]_{kn\times kn}\,,
\ee
where $F_-,F_+F,B,C$ are square matrices which commute with each other. Before the calculation of this determinant, we consider a result that support the calculation. Let‘s start from a simple example.
We consider a matrix
\be\label{matM}
	M_2=\left(\begin{array}{cc}
	A& B  \\
	C& D \end{array}\right)\, ,
\ee
in which all of A, B, C and D are matrices and commute with
each other. The determinant of $M$ can be
evaluated by, without losing generality supposing $A$ is invertible,
\be  \label{derM_2}
 \det(M_2)=\det \Big(\left(\begin{array}{cc}
	I& 0  \\
	-CA^{-1}& I \end{array}\right)
	\left(\begin{array}{cc}
	A& B  \\
	C& D \end{array}\right)\Big)=
	\det \left(\begin{array}{cc}
	A& B  \\
	& D-CA^{-1}B \end{array}\right)=  \det(AD-BC)\,.
\ee
%
%where
%\be
%    \det \left(\begin{array}{cc}
%	I& 0  \\
%	-CA^{-1}& I \end{array}\right)=1.
%\ee
The matrix $AD-BC$ in Eq.(\ref{derM_2})
will be %called ``blocked determinant matrix" of $M_2$, and
shortened as ${\rm Blodet} M_2$. Next we consider a more general situation.
We consider two matrices $m_{n\times n}$ and  $M_{2n\times 2n}$:
 \be\label{mmatr}
	m=\left[\begin{array}{ccccc}
	m_{11} & m_{12} & m_{13}&... &
	\\m_{21} & m_{22}  & m_{23}& ... &
	\\m_{31} & m_{33}  & m_{33}  & ...&
	\\ ... &  & &...  &
	\\ &  & & m_{n-1,n} &m_{nn} \end{array}\right]_{n\times n}\,,
\ee
 \be\label{mmatr}
	M=\left[\begin{array}{ccccc}
	M_{11} & M_{12} & M_{13}&... &
	\\M_{21} & M_{22}  & M_{23}& ... &
	\\M_{31} & M_{33}  & M_{33}  & ...&
	\\ ... &  & &...  &
	\\ &  & & M_{n-1,n} &M_{nn} \end{array}\right]_{kn\times kn}\,,
\ee
where all $M_{ij}$ are square matrices which commute with each other and non-degenerate if nonzero.  The determinant of $m$ can be calculated as follows
\be
	\det m=\epsilon_{i_1i_2...i_n}m_{1i_1}m_{2i_2}...m_{ni_n}\,,
\ee
we are going to show that
\be\label{detM}
	\det M=\det (\epsilon_{i_1i_2...i_n}M_{1i_1}M_{2i_2}...M_{ni_n})\,.
\ee
$m$ can be brought to upper triangular form by a similarity transformation
\be
	s^{-1}ms=\text{uptri} (m) =\left[\begin{array}{ccccc}
	\overline{m}_{11} &\overline{m}_{12}  &\overline{m}_{13} &... &\overline{m}_{1n}
	\\ & \overline{m}_{22}  &\overline{m}_{23}  &... &\overline{m}_{2n}
	\\ &    & \overline{m}_{33}  &... &\overline{m}_{3n}
	\\  &  & &... &...
	\\ &  & &  &\overline{m}_{nn} \end{array}\right]_{n\times n}\,,
\ee
where the elements of $s$ and $\text{uptri} (m) $ are functions of $m_{ij}$ by $+,-,\times,\div$ four operations. And we have
\be
	\det m= \det \text{uptri} (m)\,.
\ee
Now we define
\be\label{tilddetM}
	{\rm Blodet} M := \epsilon_{i_1i_2...i_n}M_{1i_1}M_{2i_2}...M_{ni_n}\,,
\ee
which is still a matrix but a very important step for calculation of the true determinant of $M$. $\widetilde{\det} M$ has the same algebraic structure that $\det m$ has. Since all $M_{ij}$ commute with each other and non-degenerate if nonzero, $+,-,\times,\div$ four operations are well-defined,
we have some $S$ such that
\be
	S^{-1}MS=\text{Blouptri} (M) =\left[\begin{array}{ccccc}
	\overline{M}_{11} &\overline{M}_{12}  &\overline{M}_{13} &... &\overline{M}_{1n}
	\\ & \overline{M}_{22}  &\overline{M}_{23}  &... &\overline{M}_{2n}
	\\ &    & \overline{M}_{33}  &... &\overline{M}_{3n}
	\\  &  & &... &...
	\\ &  & &  &\overline{M}_{nn} \end{array}\right]_{kn\times kn}\,,
\ee
so
\be\label{Mprod}
	{\rm Blodet} M={\rm Blodet}~  \text{Blouptri} (M)&=& \prod_i \overline{M}_{ii}\,.
\ee
Moreover, the true determinant of $M$ is computed as
\be\label{detMprod}
	\det M= \det  \prod_i \overline{M}_{ii}\,.
\ee
Combination of Eq. (\ref{Mprod})  Eq. (\ref{detMprod}) and Eq. (\ref{tilddetM}) gives
\be
    	\det M= \det  \prod_i \overline{M}_{ii}=\det {\rm Blodet} M=\det (\epsilon_{i_1i_2...i_n}M_{1i_1}M_{2i_2}...M_{ni_n})\,,
\ee
which is Eq. (\ref{detM}). So to calculate $\det M$ we can first calculate ${\rm Blodet}M$. Now we can finally compute Eq. (\ref{Dn}). Defining
\be\label{DnBlo}
	\widetilde{\bold{BloD}}_n= {\rm Blodet}\,\left[\begin{array}{ccccc}
	F_- & B & & &
	\\C & F  & B &  &
	\\0 & C & F & B&
	\\  &  & &...  &
	\\ &  & & C &F_+ \end{array}\right]_{kn\times kn}\,,
\ee
we have, from Eq. (\ref{tildn} ) and (\ref{detM}),
\be
	\sqrt{F^2-4BC}\,\widetilde{\bold{BloD}}_n&=&\Big((F_--\frac{F}{2})(F_+-\frac{F}{2})+\frac{F^2-4BC}{4}\big)\Big(R_+^{n-1}-R_-^{n-1}\Big)\nonumber
	\\
	&&+(F_-+F_+-F)\frac{\sqrt{F^2-4BC}}{2}\Big(R_+^{n-1}+R_-^{n-1}\Big)\,,
\ee
here
$$R_{\pm}=\frac{F\pm\sqrt{F^2-4BC}}{2}.$$
$\sqrt{F^2-4BC}$ are well-defined because $F, B, C$ commute with each other so they can be simultaneously diagonalized. Then we have
\be\label{dettilbD}
	\det \sqrt{F^2-4BC}\,\widetilde{\bold{D}}_n&=&\det \Big[ \Big((F_--\frac{F}{2})(F_+-\frac{F}{2})+\frac{F^2-4BC}{4}\big)\Big((\frac{F+\sqrt{F^2-4BC}}{2})^{n-1}-(\frac{F-\sqrt{F^2-4BC}}{2})^{n-1}\Big)\nonumber
	\\
	&&+(F_-+F_+-F)\frac{\sqrt{F^2-4BC}}{2}\Big((\frac{F+\sqrt{F^2-4BC}}{2})^{n-1}+(\frac{F-\sqrt{F^2-4BC}}{2})^{n-1}\Big)\Big]\,,
\ee
With the following substitution,
\be
	F\to F ({\cal E}),\, F_-\to F _-({\cal E}),\, F_+\to F_+ ({\cal E}),\, B\to -t_\bot(h_1-ih_2),\, C\to -t_\bot(h_1+ih_2)
\ee
we get the expression for ${\rm det}_n({\cal E})$ and ${\rm det}_n({\cal} E=h_0)$, which is  Eq. (\ref{dnEABC}).
%%%%%%%%%%%%%%%%%%%%%%%%%%%%%%%%%%%%
 %\begin{acknowledgments}
%\end{acknowledgments}

%%%%%%%%%%% References %%%%%%%%%%%%%%%%%%%%%%%%%
%\newcommand{\J}[4]{#1 {\bf #2} (#3) #4}
%\newcommand{\andJ}[3]{{\bf #1} (#2) #3}
%\newcommand{\AP}{Ann.\ Phys.\ (N.Y.)}
%\newcommand{\MPL}{Mod.\ Phys.\ Lett.}
%\newcommand{\NP}{Nucl.\ Phys.}
%\newcommand{\PL}{Phys.\ Lett.}
%\newcommand{\PR}{ Phys.\ Rev.}
%\newcommand{\PRL}{Phys.\ Rev.\ Lett.}
%\newcommand{\PTP}{Prog.\ Theor.\ Phys.}
%\newcommand{\hep}[1]{{ hep-th/{#1}}}
%%%%%%%%%%%%%%

%\bibliographystyle{ytphys}
%\bibliography{MultilayerHaldane2}
\bibliography{MultilayerHaldaneSSCRevised}

\end{document}